\documentclass{ws-ijmpe}

\def\PRL{\rm Phys. Rev. Lett.\,}

\def\PRC{{\rm Phys. Rev.} C\,}

\newcommand{\etal}{{\em et al.}}
\newcommand{\gevsq}{$\mathrm{GeV}^2$}

\newcommand{\qsq}{$Q^2$}

\newcommand{\gen}{$G_E^n$}
\newcommand{\gmn}{$G_M^n$}
\newcommand{\gep}{$G_E^p$}
\newcommand{\gmp}{$G_M^p$}

\begin{document}

\markboth{Kees de Jager}{The SuperBigBite Project}

\catchline{}{}{}{}{}

\title{THE SUPERBIGBITE PROJECT\\
A STUDY OF NUCLEON FORM FACTORS}

\author{\footnotesize KEES DE JAGER\\
for the JLab Hall A EMFF collaboration}

\address{Jefferson Lab, 12000 Jefferson Avenue\\
Newport News, VA 23606,
USA}

\maketitle

\begin{history}
\received{(received date)}
\revised{(revised date)}
\end{history}

\begin{abstract}
A proposed set of instrumentation, collectively referred to as the Super Bigbite 
project, is presented. Used in three different configurations it will allow measurements 
of three nucleon electromagnetic form factors \gen, \gep, and \gmn with unprecedented 
precision to \qsq-values up to three times higher than existing data.
\end{abstract}

\section{Scientific Motivation}

The study of nucleon form factors has seen an enormous revival through 
the discovery by Jones {\it et al.}\cite{gep1}, that \gep/\gmp~ drops almost linearly
with \qsq\ above a four-momentum transfer of something like 
1 GeV$^2$.  
Those results have stimulated huge amounts of theoretical activity, as evidenced 
by the nearly 500 citations of their original paper.   
One approach to explaining the data involves refined perturbative 
QCD calculations that include 
an $L=1$ component in the quark light-cone wave function\cite{BJY}.  
Also notable are relativistic constituent-quark calculations\cite{rcqm}, 
some of which even preceded the discovery by Jones \etal\  
Perhaps the most realistic model is a calculation out of Argonne by Clo\"et, 
Roberts and coworkers that uses an approach based on 
the Dyson Schwinger Equations (DSEs) together with the  Poincar\'e-covariant 
Faddeev equations\cite{clo09}.  
Here, the constituent quarks have their masses dynamically generated using 
the DSE approach.  
While still a model, the DSE/Faddeev calculation from Argonne offers a solution that
consistently incorporates both QCD-based dynamics along with dressed diquark 
degrees of freedom. Also, it is well constrained by the nucleon's static properties 
such as mass and magnetic moment.  
It is limited, however, in that there are precisely three (and for instance, 
not five) constituent quarks used as input to the calculation.  
Even so, it is reasonable to assume dominance of 
the 3-quark component of the wave function at relatively high values of \qsq.
Finally, truly {\it ab initio} calculations of form factors using lattice QCD
have been performed, some of which are extrapolated to a realistic pion-mass 
value using chiral perturbation theory\cite{lqcd}.  
While currently limited in their \qsq-reach,  it is likely that 
such calculations will reach higher \qsq-values with timescales 
consistent with the measurements proposed here.

\begin{figure}[!ht]
\unitlength 1cm
\begin{minipage}[th]{6cm}
\includegraphics[width=1.0\textwidth]{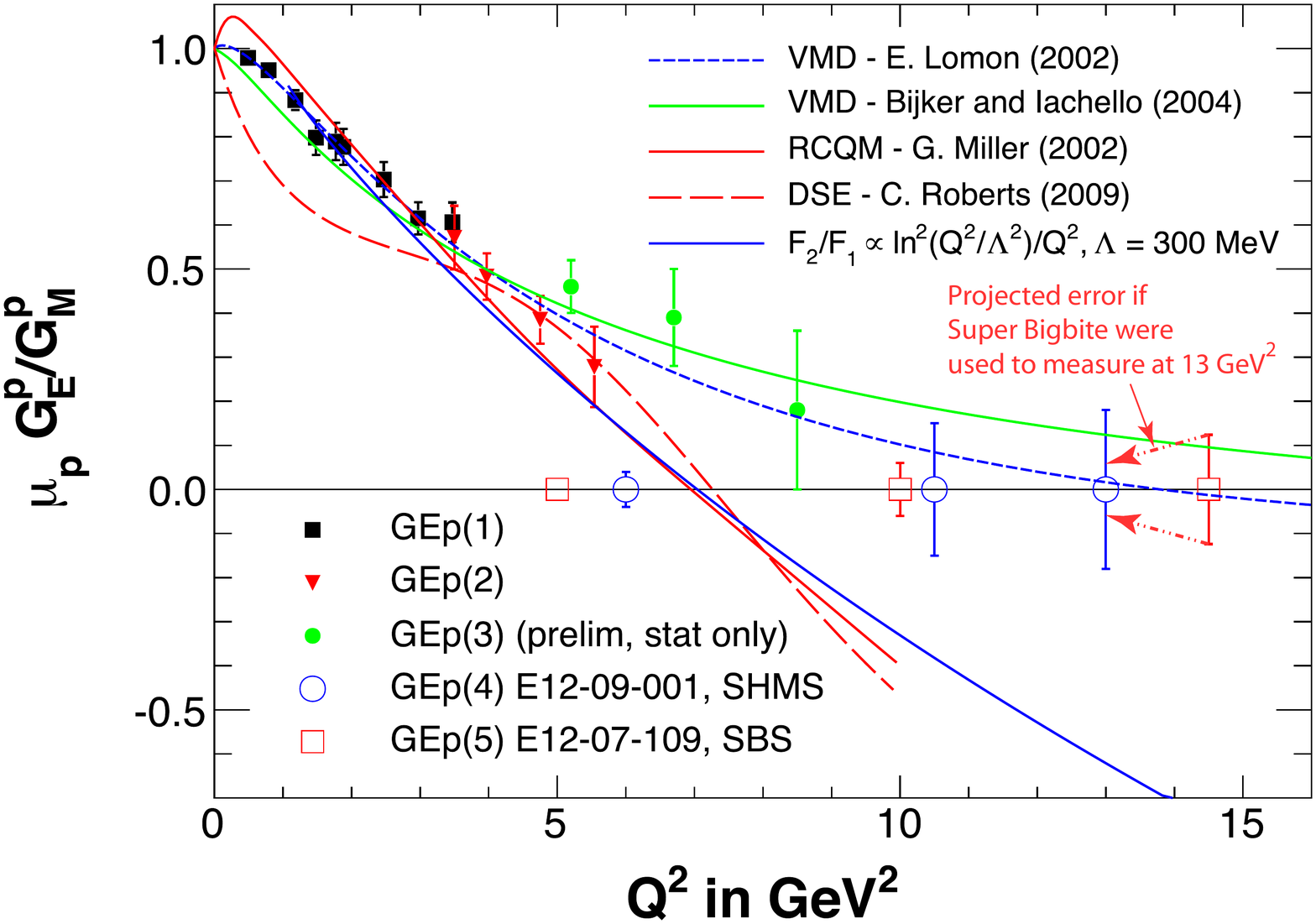}
\end{minipage}
\hskip 0.1truein
\begin{minipage}[th]{6cm}
\includegraphics[angle=0, width=1.0\textwidth]{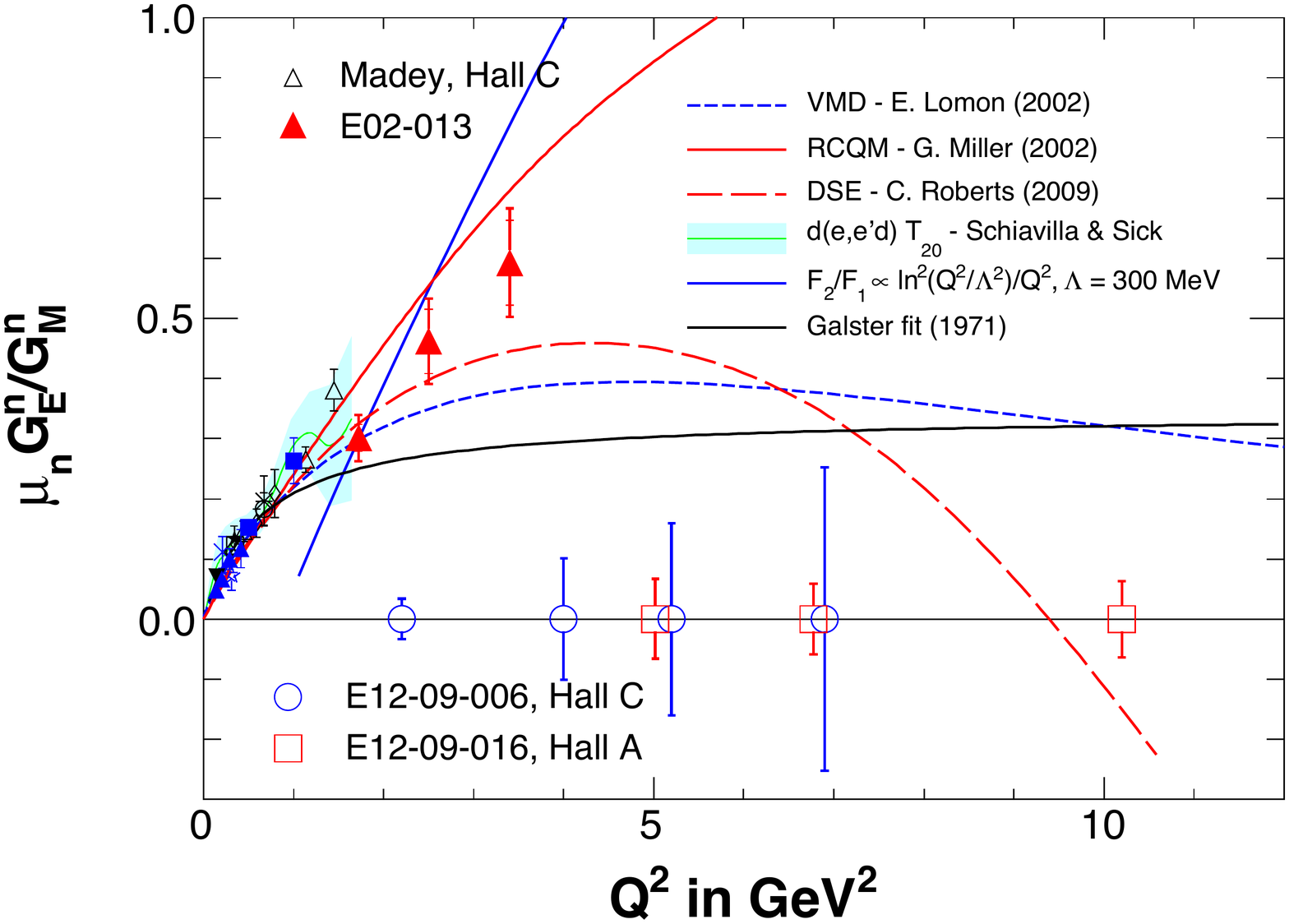}
\end{minipage}
\hfill
\caption[]{Shown are existing data and projected errors for measurements of the ratios 
of the electric and magnetic form factors of the proton (left panel) and neutron (right panel).   
The projected errors for the measurements made within the Super Bigbite project 
are shown by the open red squares.
On the left panel are shown the published results of GEp(1)\cite{gep1}, GEp(2)\cite{gep2}, 
preliminary results from GEp(3)\cite{gep3}, the proposed points of GEp(4) with SHMS 
in a 90-day run\cite{gep4} and the projected results of GEp(5) in a 60-day run\cite{gep5}.
To emphasize the larger statistical power (a factor of ten) of GEp(5),
we indicate on the figure the size of the error bars that would be achieved
using SBS to run at 13 GeV$^2$.
On the right-hand panel, for \gen/\gmn, we show published data including those due to Madey and coworkers\cite{madey}, and preliminary
results of GEn(1) (E02-013)\cite{gen1}. We also show the projected errors of GEn(2)\cite{gen2}, which is part of the Super Bigbite project, and 
E12-09-006\cite{GEn_HallC} with SHMS (open blue points).  Note that at roughly 7 GeV$^2$, the highest value of the SHMS experiment, GEn(2) achieves an error four times smaller, and with one quarter of the running time. This illustrates the $\times$50 advantage in Figure-of-Merit of GEn(2) over the SHMS experiment.
\label{fig:ratio_ff}
}
\end{figure}

Figure~\ref{fig:ratio_ff} shows existing data for \gep/\gmp\ and \gen/\gmn, 
the projected errors for several approved experiments, and the results of 
several theoretical calculations.  
The approved experiments associated with the Super Bigbite experiment are E12-07-109, 
also known as GEp(5), and E12-09-016, which will measure  \gep/\gmp\ and \gen/\gmn\ , 
respectively.  
Figure~\ref{fig:ratio_ff} makes it clear that the only way to achieve clarity in 
discriminating between theoretical explanations of the form-factor data is to measure 
the form factors with considerable precision to high values of \qsq~ in 
{\it both the proton and the neutron}.  
For example, three of the predictions shown, the relativistic constituent-quark model 
(RCQM)\cite{rcqm}, 
the DSE/Faddeev calculation\cite{clo09}, and the refined pQCD 
calculation\cite{BJY} ($F_2/F_1\propto \ln^2(Q^2/\Lambda^2)/Q^2$), 
all show \gep/\gmp\ crossing zero somewhere in the neighborhood of 
7 GeV$^2$.  
At the same time, the two VMD models\cite{lomon,iachello} show \gep/\gmp\  approaching
zero much more gradually.  
{\it In contrast}, in the case of the neutron, even by 10 GeV$^2$, the RCQM, pQCD and 
DSE/Faddeev calculations all differ wildly from one another.  
In the years following the discovery by Jones \etal, 
it is not surprising that models have evolved that explain well 
the existing proton data.  
It is also not surprising that these models diverge strongly where
there are little or no data to constrain the calculations, such as at 
higher \qsq~for the proton
or even moderate \qsq~for the neutron. In general,  higher values
of \qsq~also offer the advantage that there are simplifications
that are not present at lower values. 
For instance, the role of vector mesons is 
suppressed at higher \qsq, as are higher 
Fock states in some of the phenomenological models.  
At high \qsq\ there is increased clarity, and increased discovery potential, 
because there are generally  fewer places to hide deficiencies in a model.

Even setting aside specific predictions, there is a crucial experimental question
that is of tremendous importance.  
The linearity of the data for \gep/\gmp\ up to something like 6 GeV$^2$
is very striking.  But to what value of \qsq~will this linearity
continue?  Looking again at Fig.~\ref{fig:ratio_ff},  the preliminary
results from GEp(3)\cite{gep3}, up to values of around $\rm 8.6\,GeV^2$,
indicate that the rate of decline of \gep/\gmp\ appears to be slowing.  
It is hard to be certain, however, because the errors are relatively large, 
larger than those projected in the original GEp(3) proposal.  
By going to 14.5 GeV$^2$, the trend of \gep/\gmp\ should become clear, 
{\it but only if the data have sufficient precision}.  
To better illustrate this point, we have shown, with the dash-dotted lines, 
the relative size of the errors one could expect for \gep/\gmp\  using 
the {\it original} projected error for GEp(3) at \qsq\ = 8.6 GeV$^2$, and 
taking the Figure-of-Merit to scale as the product of the cross section $\sigma$ and the square of the analyzing power A$_y$, $\sigma \cdot A^2_{_y}$.  
We have assumed 30 days of data taking, as was assumed for the 8.6 GeV$^2$ point.  
With this scaling, it is clear that the projected errors for GEp(4) represent 
a considerable challenge in terms of beam time. In contrast,
the projected errors for GEp(5), which is based on Super Bigbite, are quite small, 
even out to 14.5 GeV$^2$.  This is because the innovative design of the
Super Bigbite approach provides fully a factor of 10 improvement
in the relevant Figure-of-Merit.  
Even with the striking behavior of \gep/\gmp\  at the values of \qsq~studied thus far, 
it is still the expectation (from pQCD) that this ratio should eventually
level off and become constant.  
The observation of a transition to this behavior would provide valuable insight, 
and it is important to have an experiment capable of achieving the required precision.

With respect to the neutron, there are strong motivations to measure
\gen/\gmn\ out to \qsq\ = 10 GeV$^2$ and even higher.  First of all,
at 10 GeV$^2$, data on the neutron will already be solidly in the regime where 
intriguing behavior has been observed with the proton.  And as pointed out
earlier, in contrast to the case with the proton, the various predictions for the neutron
all disagree strongly with one another for  \qsq\ = 10 GeV$^2$.
Also, at \qsq\ = 10 GeV$^2$,
as is shown in Fig.~1, the Argonne DSE/Faddeev calculation shows a zero crossing for
 \gen/\gmn, a feature due to the use of diquark degrees of freedom.  As mentioned above, 
the Argonne calculation, while definitely still a model, 
contains features such as the dynamical generation of mass 
that suggest progress toward an actual analytical solution.  
It is intriguing that this sophisticated calculation appears to be so 
successful, and it is desirable to test its predictions as thoroughly as possible.
With this in mind, we note that at  7 GeV$^2$, the predictions
of the VMD models and the Argonne DSE/Faddeev calculation are fairly close 
to one another. Thus, measuring well above 7 GeV$^2$ should be 
a priority.  Indeed, looking again at Fig.~1, our proposed measurement of  \gen/\gmn\ 
(E12-09-016) provides good precision all the way to 10 GeV$^2$.
In sharp contrast, the projected errors for the only competing experiment
(E12-09-006) provide little discrimination between different theories even at 7 GeV$^2$,
the highest \qsq~point for which it would provide data.  
The significant difference between E12-09-016 and E12-09-006 results from
an intrinsic Figure-of-Merit associated with the Super Bigbite approach that is roughly
a factor of 50 larger than is the case with E12-09-006.

There are, of course, many motivations for measuring the ground-state
form factors that are not illustrated in the above figure.  For instance, 
for both the proton and the neutron,  the ground-state elastic form factors 
provide stringent model-independent constraints on Generalized Parton 
Distributions (GPDs).   Thus, if we want to know the GPDs over a wide kinematic 
range, we need to study the elastic form factors over a similar range.  
The elastic form factors also provide a powerful check of lattice QCD.  
{\it Ab initio} lattice calculations of 
ground-state form factors are making impressive progress, and  the comparison
of these results with experimental measurements is extremely important.
Here, it is critical to have experimental results on {\it both} the proton and the neutron, 
and to cover the {\it full} \qsq-range being explored on the lattice. 
By the time the CEBAF upgrade is complete and the measurements described here
 have been made, it is quite likely that lattice results will 
extend up to roughly 10 GeV$^2$, again making coverage of this range for both the
proton and neutron extremely important.

\begin{figure}[!ht]
\unitlength 1cm
\begin{minipage}[th]{5cm} \hskip -0.2truein
\includegraphics[width=1.6\textwidth, angle = 0.]{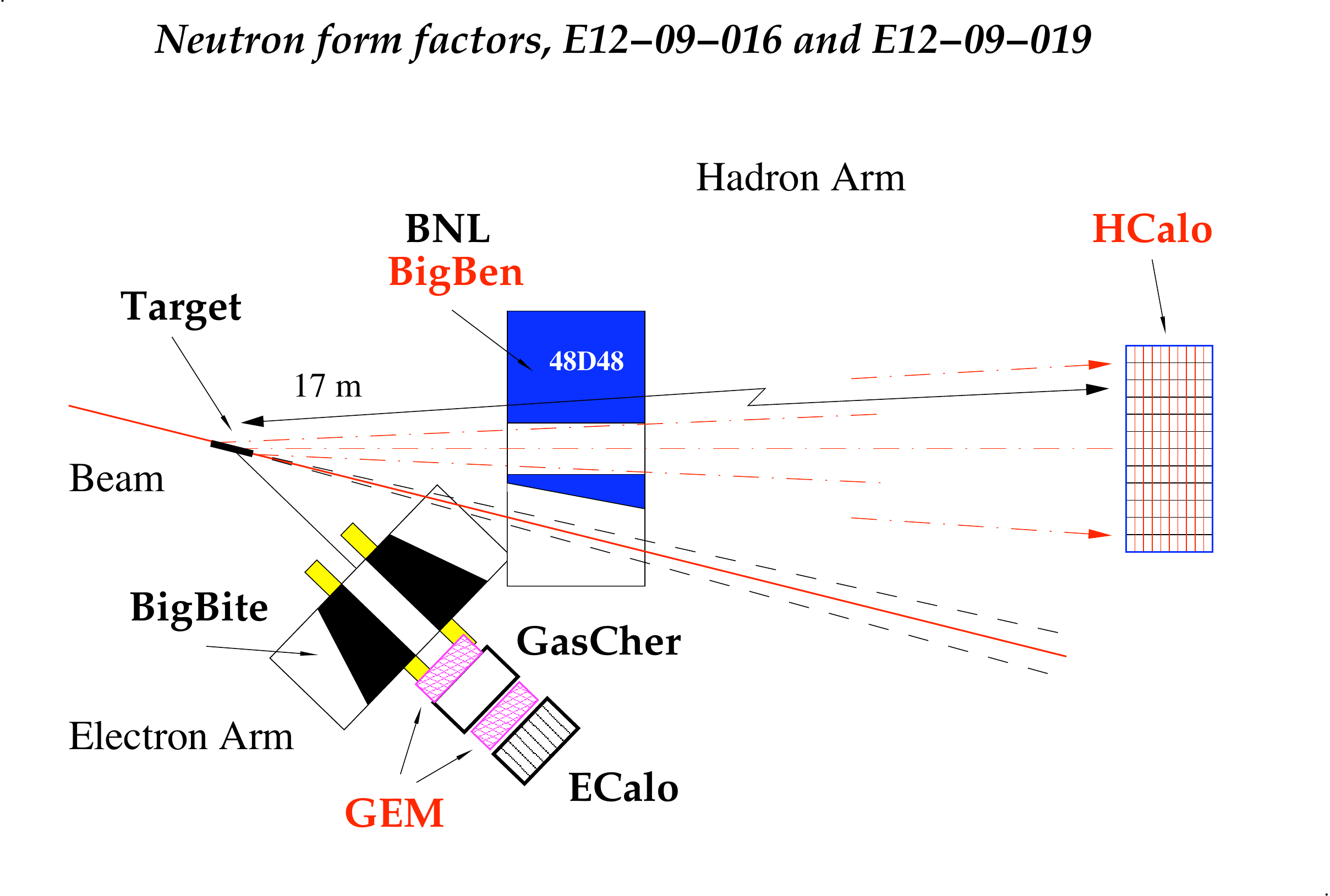}
\end{minipage}
\hfill
\begin{minipage}[th]{4cm}
\caption[]{Shown is a schematic but scaled representation of the setup that will be used for both the 
GEn(2) (E12-09-016) and the Hall A GMn experiments (E12-09-019).  While the target will be polarized $^3$He for GEn(2) and deuterium for the GMn experiment, most other components are identical.}
\end{minipage}
\label{fig:GEN-layout}
\end{figure}

The Hall A GMn experiment will determine
\gmn\ by a detailed comparison of the unpolarized elastic cross sections of
the two processes $d(e,e'\,p)n$ and $d(e,e'\,n)p$.  It will use essentially the 
same apparatus as GEn(2), with the exception that the target will be 
the Hall A liquid deuterium cryotarget.  A schematic representation of the experimental
setup is shown in Fig.~2.
We note that the GMn proposal actually included measurements up to 18.0 GeV$^2$,
something that, combined with the fully approved Hall A \gmp\ measurement 
(not part of the Super Bigbite project), would enable the reconstruction of
the individual $u$ and $d$ quark distributions with a spatial resolution of 0.05~fm.  
The EMFF collaboration plans to return to the PAC
to request an additional two weeks to push to this higher $Q^2$, as the 
difference between the $u$ and $d$ quark distributions is an exciting question
with implications for our understanding of nucleon structure in terms of QCD
degrees of freedom.

Like the other Super Bigbite project form-factor measurements, the \gmn\ measurement in Hall A
will provide excellent accuracy and reach in $Q^2$, well beyond all competing efforts.  
Considering only the portion  of the experiment that is fully approved, the GMn experiment will
require only 14 days of running.  The CLAS12 \gmn\ experiment, which like the Hall A experiment 
is approved to make measurements up to 13.5 GeV$^2$, will require 56 days
of running, and will obtain 5 times less statistics at the highest $Q^2$-point.
When considering a full set of kinematics, the Hall A \gmn\ experiment has a Figure-of-Merit
that is 30 times higher than that of the CLAS12 \gmn\ experiment.  The existing data for \gmn, together
with the projected errors for both the Hall A and the CLAS12 experiments, are shown
in Fig.~3.  While the magnetic FF of the neutron has previously been measured up to 10~GeV$^2$,
the few data that exist above 4.5 GeV$^2$ have uncertainties of about 10-20\%.
The GMn experiment in Hall A will provide sufficient accuracy to bring new understanding
to this subject, including the aforementioned decomposition of the $u$ and $d$-quark distributions.

\begin{figure}[!ht]
\unitlength 1cm
\begin{minipage}[th]{6cm}
\includegraphics[width=1.4\textwidth, angle = 0.]{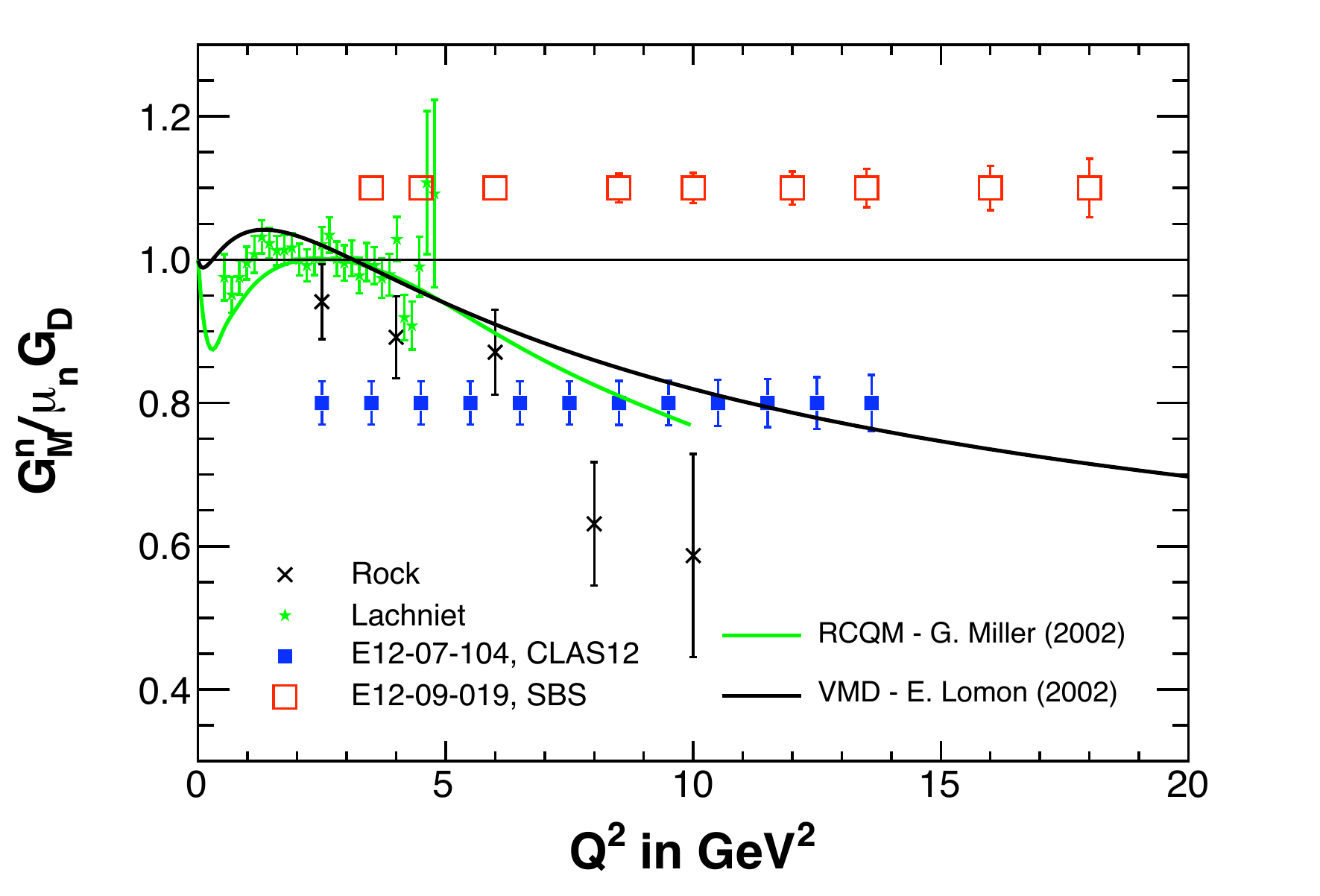}
\end{minipage}
\hfill
\begin{minipage}[th]{4cm}
\caption[]{Published \gmn~data together with the proposed data points of the
CLAS12 \gmn\ experiment (E12-07-104), and the proposed data 
of the Hall A \gmn\ experiment (E12-09-019) that will be performed using the Super Bigbite apparatus.}
\end{minipage}
\label{fig:GMN}
\end{figure}

The ratio method has been used in a number of experiments, including one at JLab in which
 \gmn\ was determined with good precision up to \qsq\ =  4.5 GeV$^2$.
When the recoil nucleon energy is above 2-3 GeV, the detection efficiency
for the neutron and the proton are quite similar, so the ratio method becomes
almost systematic free.  Like GEn(2), the Hall A GMn experiment will utilize
the BigBite spectrometer with the caveat that the trackers will be
based on GEMs instead of MWDCs.  Again like GEn(2), the Super Bigbite magnet
itself will be placed in the hadron arm, providing excellent separation between recoil
protons and recoil neutrons.  The magnet will also be turned on and off to study potential systematics.  

Historically, and even quite recently, the ground-state form factors have 
provided considerable insight into the charge and magnetization 
densities of both individual nucleons as well as more complex nuclei.  
It is quite reasonable to argue that form factors provide us with the best
``snapshots" we have of both the proton and the neutron.  At small 
values of \qsq, the Fourier transforms of the electric and magnetic form factors
can be interpreted as the charge and magnetization densities, respectively.  
Such interpretations by Hofstadter provided early understanding of the
size of the proton, and the more recent application of such reasoning
has led to the conclusion that the charge and magnetization densities of
the proton are not coincident with one another.  
The neutron charge density is positive toward its center and is surrounded 
by negative charge, features that support the notion of a proton-like 
core embedded in a negative pion cloud.  We should note, however,
that relativistic effects limit the degree to which the interpretation
of form factors as Fourier transforms of densities is correct.  
Attempts to better account for relativity in the lab frame have been
conducted by Kelly\cite{kel02}. 
Another approach, pursued by Miller, is to define densities
that can be computed on the light front\cite{mil07}.  
The aforementioned relativistic constituent-quark models 
and calculations based on the
DSE/Faddeev approach both incorporate the basic idea of pion clouds.
A nucleon with a pion cloud necessarily represents
a five-quark state, something that is suppressed at high momentum transfer 
when \qsq $\gg \Lambda_{QCD}^2$.  There is considerable
discovery potential in pushing  to higher values of \qsq~ where the 
short-distance-scale behavior of the nucleon can be revealed, and the
structure itself becomes simpler and easier to understand.  

All the arguments above underscore the importance of reaching high \qsq~while simultaneously 
maintaining high precision.  
The required beam time, however, scales roughly as $Q^{16}/E_{beam}^{2}$.  
It is thus critical to compensate for this large factor in the chosen experimental design.  
For the measurements proposed within the Super Bigbite 
project, the relevant Figure-of-Merit exceeds those of all competing experiments by factors ranging 
between 10 and 50, making the difference between meaningful and ambiguous measurements.   
These impressive capabilities are derived from using a large open-geometry dipole magnet 
for momentum analysis together with a detector package that has a direct view of the target.  
The very high rates associated with such a configuration are only tolerable because of the use 
of a GEM-based tracking system.  
A cutout in the Super Bigbite magnet also allows for 
the use of quite forward angles where the recoil nucleon needs to be detected 
with large solid angle. 
The feasibility of such an open geometry design has been unambiguously
demonstrated 
in multiple experiments using Super Bigbite's predecessor, BigBite.  
The Super Bigbite apparatus, however, based on the use of a larger magnet combined with 
GEM-based trackers, leads to a Figure-of-Merit that exceeds that of other competing efforts
 by a factor of 10 for \gep/\gmp, around 30 for \gmn,  and 50 for \gen/\gmn.  
Super Bigbite will advance the study of 
electromagnetic form factors in a dramatic fashion {\it for both the proton and the neutron}.  
Without Super Bigbite, however, as shown earlier, measurements will be largely incapable 
of discriminating between the important theoretical predictions.

The Super-Bigbite apparatus will make possible three ground-breaking
measurements of the nucleon's elastic form factors:
\begin{itemize}
\item A measurement of \gen~(E12-09-016) up to 10 GeV$^2$ using the
beam-target double-polarization technique that was approved in January of 
2009\cite{gen2}.
\item A measurement of \gep~(E12-07-109) up to 14.5 GeV$^2$ using 
the recoil-polarization technique that was approved 
by the JLab Program Advisory Committee (PAC) in August of 2007\cite{gep5}.  
\item A measurement of \gmn~(E12-09-019) up to 13.5 GeV$^2$ by determining 
the cross section ratio for the two reactions D(e,e'n) and D(e,e'p) that 
was approved by the JLab PAC in January of 2009\cite{gmn} 
(a request for an extension to 18 GeV$^2$ is planned).
\end{itemize}

We note that the above experiments will make use of the results of 
JLab E12-07-108\cite{gmp}
(not part of the Super Bigbite Project) that will measure \gmp~ up to
17.5 GeV$^2$.  E12-07-108 was approved by the JLab PAC in 2007 and 
will use the exquisite calibration of the Hall~A HRS spectrometers 
to achieve a 1-2\% absolute measurement of the electron-proton elastic 
scattering cross section.
This calibration will allow us to measure ratios of form factors rather than 
the absolute form factors themselves while still achieving our goals 
for absolute measurements.  
We thus have a plan to measure all of the ground-state electromagnetic 
form factors with sufficient accuracy and reach in \qsq~to study some 
of the most exciting questions in hadronic physics.

 \section{Instrumentation}

The proposed instrumentation, that we refer to collectively as 
the Super Bigbite Project,  includes a set of components
that will be used in three different configurations for each of 
our measurements of \gen, \gep, and \gmn, respectively.
In all cases, the design philosophy incorporates  the use of large 
open-geometry detection, high-rate-handling capability through 
the use of GEM technology, and the ability to measure at relatively 
forward angles.  
We describe our proposed instrumentation below, including the 
various configurations 
in which they will be used for our three proposed measurements.

 \vskip 0.2truein

The proposed Super Bigbite apparatus is shown in Fig.~4 
in the configuration for the GEp(5) experiment\cite{gep5}.
The key features of the apparatus are:
\begin{itemize}
\item The dipole magnet, which is an existing 48D48 used previously at BNL. 
\item Large solid angle, 10-15~times 
larger than in focusing spectrometers, such as HRS/HMS/SHMS.
\item Large momentum acceptance, from 2~GeV/$c$ at nominal field settings.
\item High luminosity capability, 
up to $8\times 10^{38}$~electron/s$\times$nucleon/cm$^2$, e.g., in GEp(5). 
\item Small scattering angle capability, down to 3.5$^\circ$.
\item Full acceptance for the long target (up to $y \approx \, \pm$20~cm). 
\item Very good angular resolution, 
$\sigma_\theta$ [mrad] = 0.14+ 1.3/$p$ [GeV/$c$].
\item Good vertex resolution, $\sigma_y \approx$ 1-2~mm.
\item Good momentum resolution, 
$\sigma_p/p = 0.0029+0.0003 \times p$ [GeV/$c$].
\item High energy trigger threshold via use of a hadron calorimeter.
\end{itemize}

\begin{figure}[!ht]
\unitlength 1cm
\begin{minipage}[th]{5.5cm} \hskip -0.2truein
\includegraphics[width=1.6\textwidth, angle = 0.]{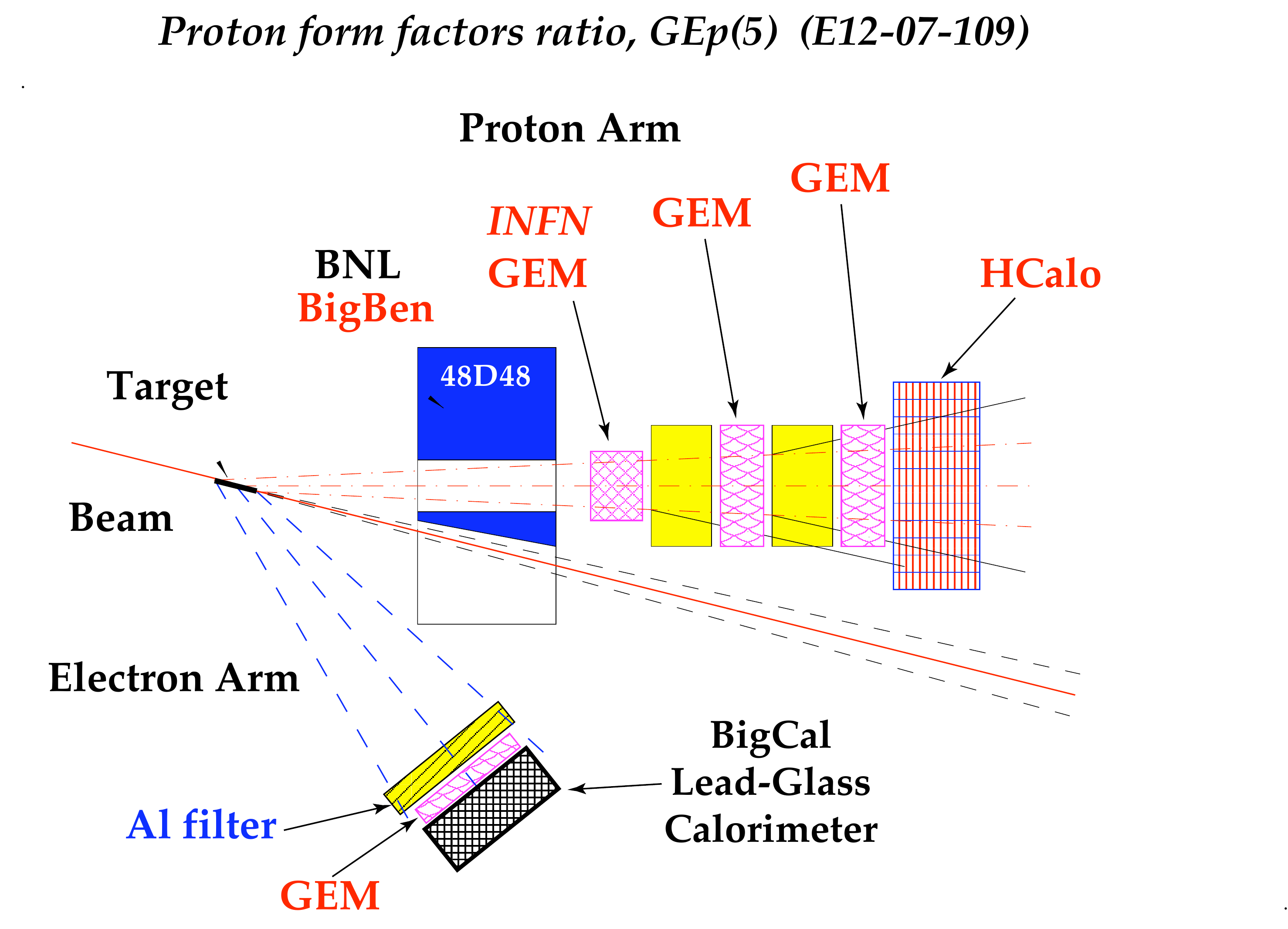}
\end{minipage}
\hfill
\begin{minipage}[th]{4.5cm}
\caption{Shown is a schematic but scale representation of the setup that will be used for GEp(5) (E12-07-109).  The proton arm (set at a central scattering angle of $12^\circ$) incorporates a double polarimeter instrumented with GEM trackers and a highly segmented hadron calorimeter.  The electron arm uses the existing ``BigCal" electromagnetic calorimeter based on lead glass.
\label{fig:GEP-layout}
}
\end{minipage}
\end{figure}

The Super-Bigbite apparatus will be used in GEp(5) as a large-acceptance 
spectrometer, the Super Bigbite spectrometer (SBS).
It will provide angular coverage up to $\sim$70~msr, 
with a detector package, capable 
of operating at the largest possible luminosity, almost
$10^{39}$~electron/s$\times$nucleon/cm$^2$.
The use of a simple dipole placed close to the target, made possible by 
a deep cut through the iron yoke of the magnet for the beam line, allows
one to achieve the large angular acceptance in the spectrometer.
The magnet deflects charged particles vertically and will be used with 
a field integral up to 2.5~T$\cdot$m.
The field in the beam line will be reduced to an acceptable level by 
specially developed magnetic shields.
The relatively small bend angle is compensated for by the high
coordinate resolution (70~$\mu$m) of the front GEM-based chambers 
resulting in a momentum resolution of $~0.5\%$ at  8~GeV/$c$
in GEp(5) with the 40-cm long LH$_2$ target.
The GEM technology solves the main challenge of this spectrometer,
the very high counting rates, allowing tracking at background rates
much higher than those expected for the experiments presented here. 
{\bf These features  combined will give  SBS at least a factor of  
10 advantage  compared with any existing or proposed spectrometer at 
Jefferson lab for nucleon form-factor measurements.}

The electron beam parameters required for the EMFF measurements, 
such as energy, polarization, intensity and size, 
are all within the 12-GeV specifications.
The GEn(2) experiment\cite{gen2} will use a polarized $^3$He target as a 
key component.
A novel concept of a convective-flow cell will allow to increase the beam
current up to 60~$\mu$A with a 60-cm-long cell.
The GEp(5) experiment will use a 40-cm long liquid hydrogen target. 
A wide vacuum snout from the scattering chamber to the
magnetic shield will allow one to avoid a direct view of the detector from 
the beam-line vacuum pipe elements, that will 
reduce the counting rate in the detector by 
a factor of 2.5.
The GMn experiment\cite{gmn} will use standard 10-cm cryogenic targets 
of hydrogen and deuterium.

\subsection{Trackers of the Super Bigbite apparatus}

There will be three trackers in SBS for the GEp(5) experiment.
The first one, FT, will be used to measure the proton momentum
and its direction before interaction with the first CH$_2$ analyzer.
This tracker with an active area of 40$\times$150~cm$^2$ will include  
six chambers, each with two-dimensional read-out and 
three GEM amplification foil planes.
The front tracker will be followed by a double polarimeter 
consisting of two trackers each preceded by a CH$_2$ analyzer.
Adding this second polarimeter increases the Figure-of-Merit 
by a factor of 1.7, equivalent to a 30\% reduction of 
the experimental statistical errors. 
The second tracker, ST, will measure the proton track after 
the proton passes through the first CH$_2$ analyzer.
The third tracker, TT, will measure the direction of the
track after the proton passes through the second CH$_2$ analyzer.
The dimensions of the ST and the TT were chosen to be 50$\times$200~cm$^2$ 
to keep the Figure-of-Merit above 90\% of the ideal value, which 
corresponds to  trackers of  unlimited size.
The ST and TT are required to have  only four chambers each 
(compared with six in the FT) due to the reduced demand on coordinate 
resolution and the lower counting rates.
Each GEM chamber will consist of 50$\times$40~cm$^2$ 
sub-sections. 

At the FT we expect high background hit rates of about 400~kHz/cm$^2$ 
(based on GEANT simulations) due to the direct view of the target. 
The background is dominated by soft photons originating from the target. 
Low-momentum charged particles are swept away by the magnet.
The rates on the second and third trackers are expected to be 130~kHz/cm$^2$
and 64~kHz/cm$^2$, respectively, dominated by soft 
electrons/positrons converted from photons in the analyzers.

\subsection {Hadron Calorimeter}

The EMFF experiments deal with very small cross sections, so
for meaningful results the luminosity should be as high as possible.
Arrangement of the trigger and the detector structure for the
high luminosity should take maximum advantage of the high energy
of the recoil nucleon.
The energy of the nucleon in the EMFF experiments ranges 
from 2 to almost 10~GeV, depending on the measurement.
In GEp(5), the large proton energy will allow 
the use of a 3-4~GeV energy threshold in the calorimeter without 
a significant loss of detection efficiency.
Such a high energy threshold leads to the  suppression of  pion triggering
and opens the possibility of using a coincidence for the DAQ trigger.

The total active area of the hadron calorimeter (HC) will be 5.5~m$^2$.
It has a good time resolution of 1.5~ns,
high granularity (15$\times$15~cm$^2$), a very good
coordinate resolution of 2~cm, and, in addition, a high
energy threshold. All these features make the HC an attractive neutron 
detector for the two neutron EMFF experiments.

Positioned at the end of the SBS detector package the HC will be used 
in the GEp(5) experiment to trigger the DAQ, in coincidence 
with the signals from the  existing  electromagnetic calorimeter, BigCal.
In addition, these two calorimeters will provide coordinate information
that will be used to locate the proton track in the FT and in the 
TT by applying kinematic constraints.
The active size of the HC in the GEp(5) experiment will be 
150$\times$300~cm$^2$, which is a little smaller than its full size
and the total number of active modules will be 200.

\subsection {Electron Arm}

In all three EMFF experiments both the (quasi) elastically scattered electron 
and the recoil nucleon will be detected.
This allows for the selection of the exclusive process, which has 
a very small cross section at high momentum transfer.

The existing Hall~A BigBite spectrometer will be used for 
the detection of  electrons in the GEn(2) and GMn experiments.
The proposed FT and two chambers 
of the  ST of SBS will be used as the tracker in the BigBite spectrometer.
This switch does not require any reconfiguration of the GEM chambers. 
The Cherenkov counter and the large  double-layer shower 
detector of the existing BigBite trigger system will be used in 
both neutron form-factor experiments. 

The existing BigCal calorimeter will be used in the GEp(5) experiment 
for detection of the scattered electron.
The calorimeter has 1744 lead-glass blocks coupled to PMTs of type FEU-84.
Two sets of blocks are used,
one 38$\times$38$\times$450~mm$^3$ and the second
slightly larger, 40$\times$40$\times$400~mm$^3$.
Blocks will be arranged in a matrix 20$\times$75,
a shape optimized for the largest acceptance at \qsq = 14.5~\gevsq.
The energy and coordinate resolutions of BigCal of about 5-7\% 
and 7~mm, respectively, for 2.5~GeV electrons satisfy the trigger 
and tracking requirements.
The calorimeter will be installed 3~m from the target 
at a central angle of 39$^\circ$ (at the largest \qsq).
It will be shielded from the target by a 20~cm Al plate
to reduce the radiation dose on the lead-glass.
The angular correlation between the scattered electron and the recoil 
proton will be measured very accurately, especially the
angle between the electron scattering plane and the proton recoil plane.
Because of the small size of the electron beam,  the angles 
of the electron and of the recoil proton can  be determined with 
a very good accuracy of $\sim$ 0.5~mrad. 
In order to achieve this,  a 1~mm coordinate accuracy  is required for the 
scattered electron.  
It will be provided by the Coordinate Detector, CD, consisting of two 
GEM-based chambers.
The sub-sections  of each chamber will be similar to those of  
the polarimeter chambers.  
The overall dimensions are 80$\times$300~cm$^2$ with 6 sections
in 2 vertical columns.

\subsection {Experimental Trigger}

The DAQ trigger will have two levels:
The first-level trigger will have a relatively
short delay, $\sim$100~ns, relative to the particle detection.
It will be generated by the electron arm.
This signal is required for operation of the GEM read-outs; despite 
their pipelined architecture, the APV25 front-end chips used in the GEMs
require a fast trigger for marking ``interesting'' samples.
Such a fast first-level trigger also eliminates the need for 
long delay lines for the conventional FASTBUS and VME electronics.
The second-level trigger will be formed using a relatively complex
coincidence logic, described below. It can have a latency of several 
microseconds. 

The trigger is a critical part of the GEp(5) experimental design.
It will have two main features: 
(1) a high energy threshold in the calorimeters in both arms 
(3-4~GeV for the proton and 2.5~GeV for the electron); and
(2) ``smart'' FPGA-based coincidence electronics that allow second-level
triggering on spatial correlations of hits in the two detector arms.
The high energy thresholds will reduce the rate in the electron arm to 60~kHz
and in the proton arm to about 1.5~MHz.
With a coincidence time window of 50~ns, forming a simple coincidence
between the two arms would result in a trigger rate of at least 5~kHz, which
is too high (or at least close to the limit) for the expected
total event size of $\sim 20$~kB.
However, the small deflection of the proton trajectory in Super Bigbite
makes it possible to use the
angular correlation between the elastic electron and the proton 
trajectories, at the trigger level, to reduce the coincidence   
trigger rate further by a factor of five.
In designing our second-level triggering system, it is important to note
that  real (not noise) calorimeter hits always cause several adjacent blocks to fire. 
We therefore construct what we call ``macroblocks", consisting of the sums of the
individual signals from several nearby blocks.  In HCAL, the size of the macroblocks
is  $4\times 4$ (we call this a ``sum-16''), and in the ECAL the size is $8\times 4$ (``sum-32''). 
The different macroblocks overlap each other, such that each individual block contributes 
to the sums associated with several different (overlapping) macroblocks.  In this way, 
at least one macroblock will always contain
a large fraction of the deposited energy associated with a real hit.

In the GMn and the GEn(2) experiments the gas Cherenkov counter and 
the two-layer shower calorimeter of the BigBite spectrometer provide
an efficient on-line selection of the electrons so
that the DAQ trigger rate will be at the level of 2--3~kHz already
without requiring an on-line coincidence with the hadron arm.

\section*{Summary}

A funding proposal for the SuperBigBite project has been submitted to DOE,
and further details about the project can be found in that proposal\cite{DOEprop}.
If the Super Bigbite project is completed in the proposed schedule, 
one of the top priorities of the JLab upgrade, 
the study of the ground-state electromagnetic 
form factors, could be achieved in the early years following commissioning.   
Studies have indicated that the Super Bigbite instrumentation provides a 
great potential for other experiments, such as studies of the neutron spin
asymmetry $A_1^n$ through inclusive scattering and of the neutron Collins 
and Sivers functions through semi-inclusive pion and kaon production off
a transversely polarized $^3$He target.

\section*{Acknowledgements}

The author gratefully recognizes the strong support provided by Gordon Cates and 
Bogdan Wojtsekhowski in preparing this manuscript. This work was supported by DOE contract DE-AC05-06OR23177, under which Jefferson Science Associates, LLC, operates the Thomas Jefferson National Accelerator Facility. 

\appendix

\end{document}